\documentclass[12pt]{iopart}

\usepackage{graphicx}

\newcommand{\be}{\begin{equation}}
\newcommand{\ee}{\end{equation}}
\newcommand{\simless}{\lower.5ex\hbox{$\; \buildrel < \over \sim\;$}}
\newcommand{\simgreat}{\lower.5ex\hbox{$\; \buildrel > \over \sim\;$}} 
\newcommand{\mion}{{ m_{\rm ion} }} 
\newcommand{\mbar}{{ \langle m \rangle }} 
\newcommand{\rhocen}{{ \rho_{\rm c}}} 
\newcommand{\thetacen}{{ \Theta_{\rm c} }} 
\newcommand{\pcent}{{ P_{\rm c} }}
\newcommand{\tcent}{{ T_{\rm c} }}
\newcommand{\tig}{{ T_{\rm nuc} }} 
\newcommand{\tstar}{{ t_{\ast {\rm min}} }}
\newcommand{\kapem}{{ \kappa_{\rm em} }} 
\newcommand{\lmax}{{ L_{\ast {\rm max}} }} 
\newcommand{\mchan}{{ M_{\rm ch} }} 
\newcommand{\konstant}{{ K_{\rm dp} }} 
\newcommand{\rgas}{{ {\cal R} }} 
\newcommand{\mzero}{{ \mu_0 }} 
\newcommand{\conlum}{{ {\cal C} }} 
\newcommand{\yield}{{ \langle \Delta E \rangle }} 
\newcommand{\mbh}{{ M_{\rm bh}}}

\newcommand{\mdm}{{ m_{\rm d} }}
\newcommand{\bcon}{{ f_{\rm g} }} 
\newcommand{\nlife}{{ N_{\rm bio} }} 
\newcommand{\lmin}{{ L_{\rm min} }} 

\begin{document} 

\title{Stars In Other Universes: Stellar structure with different fundamental constants} 


\author{Fred C. Adams} 

\address{\obeylines 
Michigan Center for Theoretical Physics, Department of Physics, 
University of Michigan, Ann Arbor, MI 48109} 

\ead{fca@umich.edu} 

\begin{abstract}     
Motivated by the possible existence of other universes, with possible
variations in the laws of physics, this paper explores the parameter
space of fundamental constants that allows for the existence of stars.
To make this problem tractable, we develop a semi-analytical stellar
structure model that allows for physical understanding of these stars
with unconventional parameters, as well as a means to survey the
relevant parameter space.  In this work, the most important quantities
that determine stellar properties --- and are allowed to vary --- are
the gravitational constant $G$, the fine structure constant $\alpha$,
and a composite parameter $\conlum$ that determines nuclear reaction
rates.  Working within this model, we delineate the portion of
parameter space that allows for the existence of stars.  Our main
finding is that a sizable fraction of the parameter space (roughly one
fourth) provides the values necessary for stellar objects to operate
through sustained nuclear fusion.  As a result, the set of parameters
necessary to support stars are not particularly rare. In addition, we
briefly consider the possibility that unconventional stars (e.g.,
black holes, dark matter stars) play the role filled by stars in our
universe and constrain the allowed parameter space.
\end{abstract} 



\maketitle

\section{Introduction} 

The current picture of inflationary cosmology allows for, and even
predicts, the existence of an infinite number of space-time regions
sometimes called pocket universes \cite{guth, rees, vilenkin}.  In
many scenarios, these separate universes could potentially have
different versions of the laws of physics, e.g., different values for
the fundamental constants of nature. Motivated by this possibility,
this paper considers the question of whether or not these hypothetical
universes can support stars, i.e., long-lived hydrostatically
supported stellar bodies that generate energy through (generalized)
nuclear processes.  Toward this end, this paper develops a simplified
stellar model that allows for an exploration of stellar structure with
different values of the fundamental parameters that determine stellar
properties. We then use this model to delineate the parameter space
that allows for the existence of stars.

A great deal of previous work has considered the possibility of
different values of the fundamental constants in alternate universes,
or, in a related context, why the values of the constants have their
observed values in our universe (e.g., \cite{carr, bartip}).  More
recent papers have identified a large number of possible constants
that could, in principle, vary from universe to universe.  Different
authors generally consider differing numbers of constants, however,
with representative cases including 31 parameters \cite{tegmark} and
20 parameters \cite{hogan}. These papers generally adopt a global
approach (see also \cite{wilczek, tegold, aguirre}), in that they
consider a wide variety of astronomical phenomena in these universes,
including galaxy formation, star formation, stellar structure, and
biology.  This paper adopts a different approach by focusing on the
particular issue of stars and stellar structure in alternate
universes; this strategy allows for the question of the existence of
stars to be considered in greater depth.

Unlike many previous efforts, this paper constrains only the
particular constants of nature that determine the characteristics of
stars.  Furthermore, as shown below, stellar structure depends on
relatively few constants, some of them composite, rather than on large
numbers of more fundamental parameters. More specifically, the most
important quantities that directly determine stellar structure are the
gravitational constant $G$, the fine structure constant $\alpha$, and
a composite parameter $\conlum$ that determines nuclear reaction
rates. This latter parameter thus depends in a complicated manner on
the strong and weak nuclear forces, as well as the particle masses. We
thus perform our analysis in terms of this $(\alpha, G, \conlum)$
parameter space.

The goal of this work is thus relatively modest.  Given the limited
parameter space outlined above, this paper seeks to delineate the
portions of it that allow for the existence of stars. In this context,
stars are defined to be self-gravitating objects that are stable,
long-lived, and actively generate energy through nuclear processes.
Within the scope of this paper, however, we construct a more detailed
model of stellar structure than those used in previous studies of
alternate universes. On the other hand, we want to retain a (mostly)
analytic model. Toward this end, we take the physical structure of the
stars to be polytropes. This approach allows for stellar models of
reasonable accuracy; although it requires the numerical solution of
the Lane-Emden equation, the numerically determined quantities can be
written in terms of dimensionless parameters of order unity, so that
one can obtain analytic expressions that show how the stellar
properties depend on the input parameters of the problem. Given this
stellar structure model, and the reduced $(\alpha, G, \conlum)$
parameter space outlined above, finding the region of parameter space
that allows for the existence of stars becomes a well-defined problem.

As is well known, and as we re-derive below, both the minimum stellar
mass and the maximum stellar mass have the same dependence on
fundamental constants that carry dimensions \cite{phil}. More
specifically, both the minimum and maximum mass can be written in
terms of the fundamental stellar mass scale $M_0$ defined according to
\be 
M_0 = \alpha_G^{-3/2} m_P = \left( {\hbar c \over G} 
\right)^{3/2} m_P^{-2} \, \approx \, 
3.7 \times 10^{33} g \approx 1.85 M_\odot \, , 
\label{mscale} 
\ee 
where $\alpha_G$ is the gravitational fine structure constant, 
\be
\alpha_G = {G m_P^2 \over \hbar c} \approx 6 \times 10^{-39} \, , 
\label{alphag} 
\ee 
where $m_P$ is the mass of the proton.  As expected, the mass scale
can be written as a dimensionless quantity ($\alpha_G^{-3/2}$) times 
the proton mass; the appropriate value of the exponent (--3/2) in 
this relation is derived below. The mass scale $M_0$ determines 
the allowed range of masses in any universe. 

In conventional star formation, our Galaxy (and others) produces stars
with masses in the approximate range $0.08 \le M_\ast/M_\odot \le
100$, which corresponds to the range $0.04 \le M_\ast / M_0 \le 50$.
One of the key questions of star formation theory is to understand, in
detail, how and why galaxies produce a particular spectrum of stellar
masses (the stellar initial mass function, or IMF) over this range
\cite{sal}.  Given the relative rarity of high mass stars, the 
vast majority of the stellar population lies within a factor of
$\sim10$ of the fundamental mass scale $M_0$. For completeness we note
that the star formation process does not involve thermonuclear fusion,
so that the mass scale of the hydrogen burning limit (at 0.08
$M_\odot$) does not enter into the process. As a result, many objects
with somewhat smaller masses -- brown dwarfs -- are also produced. One
of the objectives of this paper is to understand how the range of
possible stellar masses changes with differing values of the
fundamental constants of nature.

This paper is organized as follows. We construct a polytropic model
for stellar structure in \S 2, and identify the relevant input
parameters that determine stellar characteristics. Working within this
stellar model, we constrain the values of the stellar input parameters
in \S 3; in particular, we delineate the portion of parameter space
that allow for the existence of stars. Even in universes that do not
support conventional stars, those generating energy via nuclear
fusion, it remain possible for unconventional stars to play the same
role.  These objects are briefly considered in \S 4 and include black
holes, dark matter stars, and degenerate baryonic stars that generate
energy via dark matter capture and annihilation. Finally, we conclude
in \S 5 with a summary of our results and a discussion of its
limitations, including an outline for possible future work. 

\section{Stellar Structure Models} 

In general, the construction of stellar structure models requires the
specification and solution of four coupled differential equations,
i.e., force balance (hydrostatic equilibrium), conservation of mass,
heat transport, and energy generation. This set of equations is
augmented by an equation of state, the form of the stellar opacity,
and the nuclear reaction rates. In this section we construct a
polytropic model of stellar structure. The goal is to make the model
detailed enough to capture the essential physics and simple enough to
allow (mostly) analytic results, which in turn show how different
values of the fundamental constants affect the results.  Throughout
this treatment, we will begin with standard results from stellar
structure theory \cite{phil, kip, hansen} and generalize to allow for
different stellar input parameters.

\subsection{Hydrostatic Equilibrium Structures}

\begin{figure}
\resizebox{\hsize}{!}{\includegraphics[]{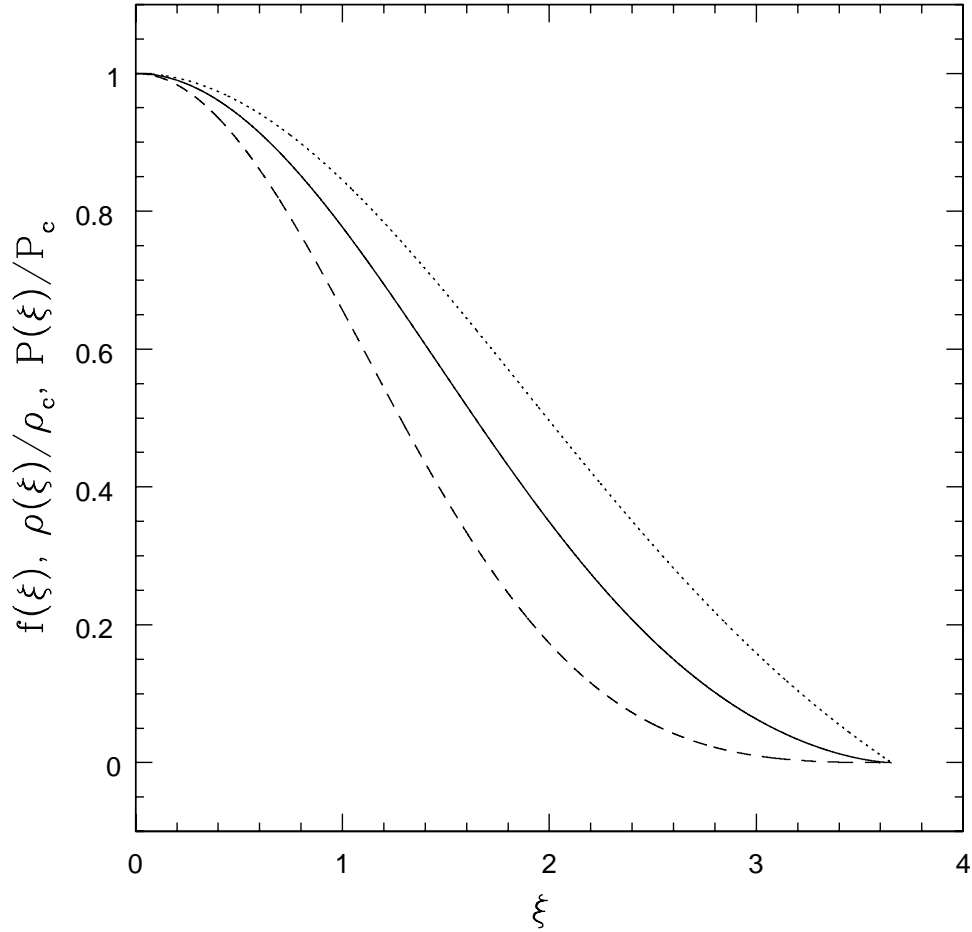}}
\caption{Density, pressure, and temperature distributions for
$n=3/2$ polytrope.  The solid curve shows the density profile
$\rho(\xi) / \rhocen$, the dashed curve shows the pressure profile 
$P(\xi) / \pcent$, and the dotted curve shows the temperature 
profile $f(\xi) = T(\xi)/\tcent$. For a polytrope, the variables 
are related through the expressions $P \propto \rho^{1 + 1/n}$ and 
$\rho \propto f^n$. }  
\label{fig:polytrope} 
\end{figure}

In this case, we will use a polytropic equation of state and thereby
replace the force balance and mass conservation equations with the
Lane-Emden equation.  The equation of state thus takes the form
\be
P = K \rho^\Gamma \qquad {\rm where} \qquad 
\Gamma = 1 + {1 \over n} \, , 
\label{polytrope} 
\ee 
where the second equation defines the polytropic index $n$.  Note that
low mass stars and degenerate stars have polytropic index $n$ = 3/2,
whereas high mass stars, with substantial radiation pressure in their
interiors, have index $n \to 3$. As a result, the index is slowly
varying over the range of possible stellar masses.  Following standard
methods \cite{chandra, phil, kip, hansen}, we define
\be 
\xi \equiv {r \over R} , \qquad \rho = \rhocen f^n , \qquad 
{\rm and} \qquad R^2 = {K \Gamma \over (\Gamma - 1) 
4 \pi G \rhocen^{2 - \Gamma} } \, , 
\label{rdef} 
\ee 
so that the dimensionless equation for the hydrostatic 
structure of the star becomes 
\be 
{d \over d\xi} \left( \xi^2 {d f \over d \xi} \right) 
+ \xi^2 f^n = 0 \, . 
\label{laneemden} 
\ee
Here, the parameter $\rhocen$ is the central density (in physical
units) so that $f^n(\xi)$ is the dimensionless density distribution.
For a given polytropic index $n$ (or a given $\Gamma$), equation
(\ref{laneemden}) thus specifies the density profile up to the
constants $\rhocen$ and $R$. Note that once the density is determined,
the pressure is specified via the equation of state (\ref{polytrope}).
Further, in the stellar regime, the star obeys the ideal gas law so
that the temperature is given by $T = P / (\rgas \rho)$, with $\rgas =
k/ \mbar$; the function $f(\xi)$ thus represents the dimensionless
temperature profile of the star.  Integration of equation
(\ref{laneemden}) outwards, subject to the boundary conditions $f = 1$
and $df/d\xi = 0$ at $\xi$ = 0, then determines the position of the
outer boundary of the star, i.e., the value $\xi_\ast$ where
$f(\xi_\ast)$ = 0.  As a result, the stellar radius is given by
\be
R_\ast = R \xi_\ast \, . 
\ee 

The physical structure of the star is thus specified up to the
constants $\rhocen$ and $R$. These parameters are not independent for
a given stellar mass; instead, they are related via the constraint
\be 
M_\ast = 4 \pi R^3 \rhocen \int_0^{\xi_\ast} \xi^2 
f^n (\xi) d\xi \, \equiv 4 \pi R^3 \rhocen \mzero \, , 
\label{mzero} 
\ee 
where the final equality defines the dimensionless quantity $\mzero$,
which is of order unity and depends only on the polytropic index $n$.

\subsection{Nuclear Reactions} 

The next step is to estimate how the nuclear ignition temperature
depends on more fundamental parameters of physics.  Thermonuclear
fusion generally depends on three physical variables: the temperature
$T$, the Gamow energy $E_G$, and the nuclear fusion factor $S(E)$.
The Gamow energy is given by
\be
E_G = (\pi \alpha Z_1 Z_2)^2 {2 m_1 m_2 \over m_1 + m_2} c^2 \, = 
(\pi \alpha Z_1 Z_2)^2 2 m_R c^2 \, , 
\label{gamow} 
\ee 
where $m_j$ are the masses of the nuclei, $Z_j$ are their charge 
(in units of $e$), and where the second equality defines the reduced
mass. For the case of two protons, $E_G$ = 493 keV. The parameter 
$\alpha$ is the usual (electromagnetic) fine structure constant 
\be
\alpha = {e^2 \over \hbar c} \approx {1 \over 137} \, , 
\label{alphae} 
\ee 
where the numerical value applies to our universe. Thus, the Gamow
energy, which sets the degree of Coulomb barrier penetration, is
determined by the strength of the electromagnetic force (through
$\alpha$). The strength of the strong and weak nuclear forces enter 
into the problem by setting the nuclear fusion factor $S(E)$, which 
in turn sets the interaction cross section according to 
\be
\sigma(E) = { S(E) \over E} \exp \left[ - 
\left( {E_G \over E} \right)^{1/2} \right] \, , 
\label{sigma} 
\ee 
where $E$ is the energy of the interacting nuclei. The temperature at
the center of the star determines the distribution of $E$. Under most
circumstances in ordinary stars, the cross section has the approximate
dependence $\sigma \propto 1/E$ so that the nuclear fusion factor
$S(E)$ is a slowly varying function of energy. This dependence arises
when the cross section is proportional to the square of the de Broglie
wavelength, so that $\sigma \sim \lambda^2 \sim (h/p)^2 \sim h^2/(2mE)$; 
this relation holds when the nuclei are in the realm of non-relativistic 
quantum mechanics.

The nuclei generally have a thermal distribution of energy so that 
\be 
\langle \sigma v \rangle = \left( {8 \over \pi m_R} \right)^{1/2} 
\left( {1 \over k T} \right)^{3/2} \int_0^\infty  \sigma(E) 
\exp \left[ - E/kT \right] E dE \, . 
\label{int} 
\ee
As a result, the effectiveness of nuclear reactions is controlled by
an exponential factor $\exp [- \Phi ]$, where the function $\Phi$ has
contributions from the cross section and the thermal distribution,
i.e.,
\be 
\Phi = {E \over kT} + \left( {E_G \over E} \right)^{1/2} \, . 
\ee
The integral in equation (\ref{int}) is dominated by energies near
the minimum of $\Phi$, where $E = E_0 = E_G^{1/3} (kT/2)^{2/3}$, and 
where the function takes the value
\be 
\Phi_0 = 3 \left( {E_G \over 4 k T} \right)^{1/3} \, .
\ee
If we approximate the integral using Laplace's method \cite{bh}, the
reaction rate $R_{12}$ for two nuclear species with number densities
$n_1$ and $n_2$ can be written in the form
\be 
R_{12} = n_1 n_2 {8 \over \sqrt{3} \pi \alpha Z_1 Z_2 m_R c} 
S(E_0) \Theta^2 \exp [-3 \Theta] \, , 
\ee 
where we have defined 
\be
\Theta \equiv \left( {E_G \over 4 k T} \right)^{1/3}  \, . 
\ee

\subsection{Stellar Luminosity and Energy Transport} 

The luminosity of the star is determined through the equation 
\be
{d L \over d r} = 4 \pi r^2 \varepsilon(r) \, , 
\ee 
$\varepsilon$ is the luminosity density, i.e., the 
power generated per unit volume. This quantity can be 
written in terms of the nuclear reaction rates via 
\be 
\varepsilon(r) = \conlum \rho^2 \Theta^2 
\exp[-3\Theta] \, , 
\ee 
where $\Theta$ is defined above, and where 
\be
\conlum = { \yield R_{12} \over \rho^2 \Theta^2} \exp[3\Theta] = 
{8 \yield S(E_0) \over \sqrt{3} \pi \alpha m_1 m_2 Z_1 Z_2 m_R c} \, \, , 
\ee 
where $\yield$ is the mean energy generated per nuclear reaction. In 
our universe $\conlum \approx 2 \times 10^4$ cm$^5$ s$^{-3}$ g$^{-1}$
for proton-proton fusion under typical stellar conditions.

The total stellar luminosity is given by the integral
\be
L_\ast = \conlum 4 \pi R^3 \rhocen^2 \int_0^{\xi_\ast} 
f^{2n} \xi^2 \Theta^2 \exp[-3\Theta] d\xi \, \equiv 
\conlum 4 \pi R^3 \rhocen^2 I(\thetacen) \, , 
\label{lumintegral} 
\ee 
where the second equality defines $I(\thetacen)$, and where
$\thetacen$ = $\Theta(\xi=0)$ = $(E_G / 4 k \tcent)^{1/3}$. 
Note that for a given polytrope, the integral is specified 
up to the constant $\thetacen$: $T = \tcent f(\xi)$, 
$\Theta = \thetacen f^{-1/3} (\xi)$. 

At this point, the definition of equation (\ref{rdef}), the mass
integral constraint (\ref{mzero}), and the luminosity integral
(\ref{lumintegral}) provide us with three equations for four unknowns:
the radial scale $R$, the central density $\rhocen$, the total
luminosity $L_\ast$, and the coefficient $K$ in the equation of state.
Notice that if the star is degenerate, then the coefficient $K$ is
specified by quantum mechanics, $\Gamma$ = 5/3, and one could solve
the first two of these equations for $R$ and $\rhocen$, thereby
determining the physical structure of the star. Note that the quantum
mechanical value of $K$ represents the minimum possible value.  If the
star is not degenerate, but rather obeys the ideal gas law, then the
central temperature is related to the central density through $\rgas
\tcent = K \rhocen^{1/n}$, so that $\tcent$ does not represent a new
unknown, and the stellar luminosity $L_\ast$ is the only new variable
introduced by luminosity equation (\ref{lumintegral}). 

For ordinary stars, one needs to use the fourth equation of stellar
structure to finish the calculation. In the case of radiative stars,
the energy transport equation takes the form
\be
T^3 {dT \over dr} = - {3 \rho \kappa \over 4 a c} 
{L(r) \over 4 \pi r^2} \, , 
\label{transport} 
\ee 
where $\kappa$ is the opacity. In the spirit of this paper, we want to
obtain a simplified set of stellar structure models to consider the
effects of varying constants. As a result, we make the following
approximation. The opacity $\kappa$ generally follows Kramer's law so
that $\kappa \sim \rho T^{-7/2}$. For the case of polytropic equations
of state, we find that $\kappa \rho \sim \rho^{2 - 7/2n}$.  For the
particular case $n = 7/4$, the product $\kappa \rho$ is strictly
constant. For other values of the polytropic index, the quantity
$\kappa \rho$ is slowly varying. As a result, we assume $\kappa \rho$
= $\kappa_0 \rhocen$ = {\sl constant} for purposes of solving the
energy transport equation (\ref{transport}). This ansatz implies that
\be 
L_\ast \int_0^{\xi_\ast} {\ell(\xi) \over \xi^2} d\xi = a \tcent^4 
{4 \pi c \over 3 \rhocen \kappa_0} R \, , 
\label{esolve} 
\ee
where we have defined $\ell(\xi) \equiv L(\xi)/L_\ast$. The full
expression for $\ell(\xi)$ is given by the integral in equation
(\ref{lumintegral}). For purposes of solving equation (\ref{esolve}),
however, we make a further simplification: We assume that the integrand
of equation (\ref{lumintegral}) is sharply peaked toward the center of
the star, and that the nuclear reaction rates depend on a power-law
function of temperature. Consistency then demands that the power-law
index is $\thetacen$. Further, the temperature can be modeled as an
exponentially decaying function near the center of the star so that $T
\sim \exp[- \beta \xi]$.  The expression for $\ell(\xi)$ then becomes
\be 
\ell(\xi) = {1 \over 2} \int_0^{x_{end}} x^2 {\rm e}^{-x} dx 
\, \qquad {\rm where} \qquad x_{end} = \beta \thetacen \xi \, . 
\label{elldef} 
\ee 
Using this expression for $\ell(\xi)$ in the integral of equation 
(\ref{esolve}), we can write the luminosity in the form 
\be 
L_\ast = a \tcent^4 {4 \pi c \over 3 \rhocen \kappa_0} 
{R \over \beta \thetacen} \, . 
\label{transolve} 
\ee

\subsection{Stellar Structure Solutions}  

With the solution (\ref{transolve}) to the energy transport equation,
we now have four equations and four unknowns. After some algebra, we
obtain the following equation for the central temperature
\be 
\thetacen I(\thetacen) \tcent^3 = 
{ (4 \pi)^3 a c \over 3 \beta \kappa_0 \conlum } 
\left( {M_\ast \over \mzero} \right)^4 
\left( {G \over (n + 1) \rgas } \right)^7 \, , 
\ee 
or, alternately, 
\be 
I(\thetacen) \thetacen^{-8} = {2^{12} \pi^5 \over 45} 
{1 \over \beta \kappa_0 \conlum E_G^3 \hbar^3 c^2} 
\left( {M_\ast \over \mzero} \right)^4 
\left( {G \mbar \over (n + 1) } \right)^7 \, . 
\label{tcsolution} 
\ee 
The right hand side of the equation is thus a dimensionless quantity.
Further, the quantities $\mzero$ and $\beta$ are dimensionless
measures of the mass and luminosity integrals over the star,
respectively; they are expected to be of order unity and to be roughly
constant from star to star (and from universe to universe). The
remaining constants are fundamental. Note that for typical values of 
the parameters in our universe, the right hand side of this equation 
is approximately $10^{-9}$. 

With the central temperature $\tcent$, or equivalently, $\thetacen$,
determined through equation (\ref{tcsolution}), we can find
expressions for the remaining stellar parameters. The radius is given
by 
\be
R_\ast = {G M_\ast \mbar \over k \tcent} {\xi_\ast \over (n+1) \mzero} \, ,
\ee 
and the luminosity is given by 
\be
L_\ast = {16 \pi^4 \over 15} {1 \over \hbar^3 c^2 \beta \kappa_0 \thetacen} 
\left( {M_\ast \over \mzero} \right)^3 \left( 
{G \mbar \over n + 1 } \right)^4 \, . 
\label{lstarsolve} 
\ee 
The photospheric temperature is then determined from the usual outer 
boundary condition so that 
\be
T_\ast = \left( {L_\ast \over 4 \pi R_\ast^2 \sigma} \right)^{1/4} \, . 
\label{tphoto} 
\ee 

\begin{figure} 
\resizebox{\hsize}{!}{\includegraphics[]{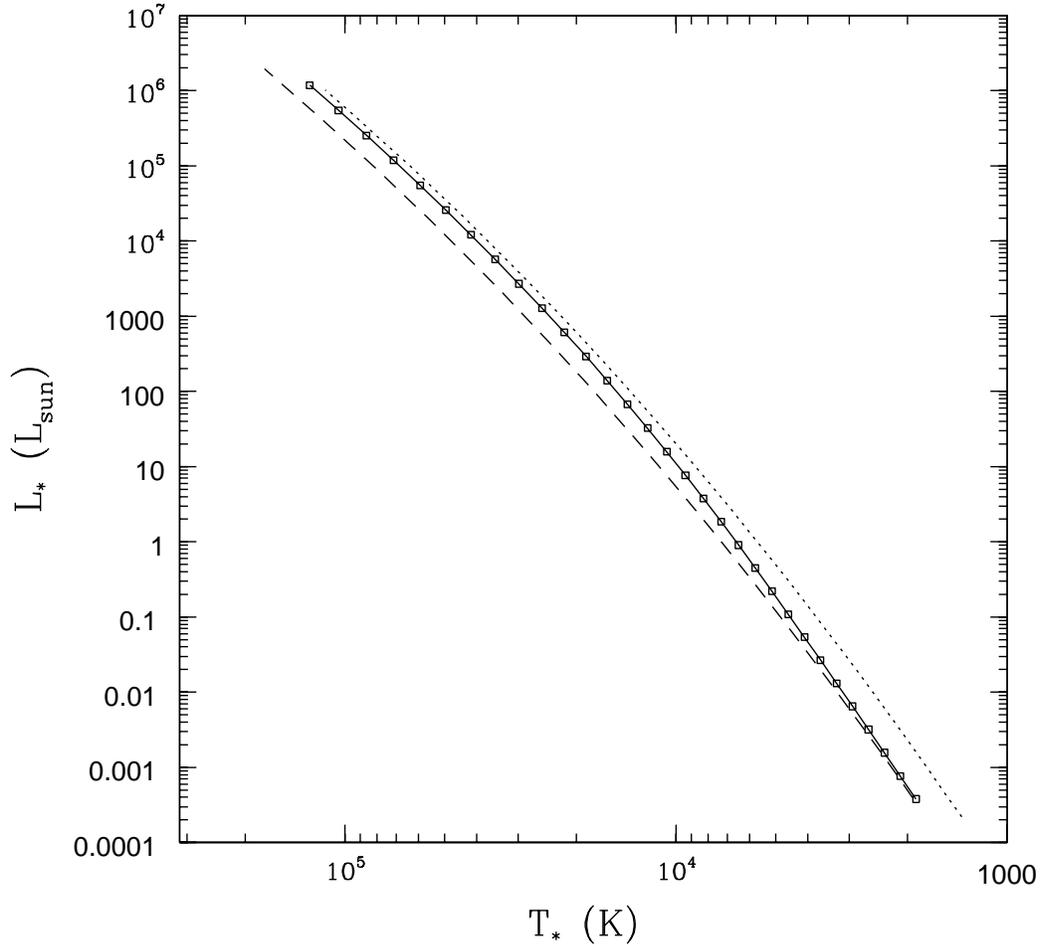}}
\caption{H-R Diagram showing the main sequence for polytropic
stellar model using standard values of the parameters, i.e., those in
our universe. The three cases shown here correspond to the main
sequence for an $n$ = 3/2 polytrope (lower dashed curve), an $n$ = 3
polytrope (upper dotted curve), and a model that smoothly varies from
$n$ = 3/2 at low masses to $n$ = 3 at high masses (solid curve marked
by symbols). }  
\label{fig:hrdiagram} 
\end{figure}

\begin{figure} 
\resizebox{\hsize}{!}{\includegraphics[]{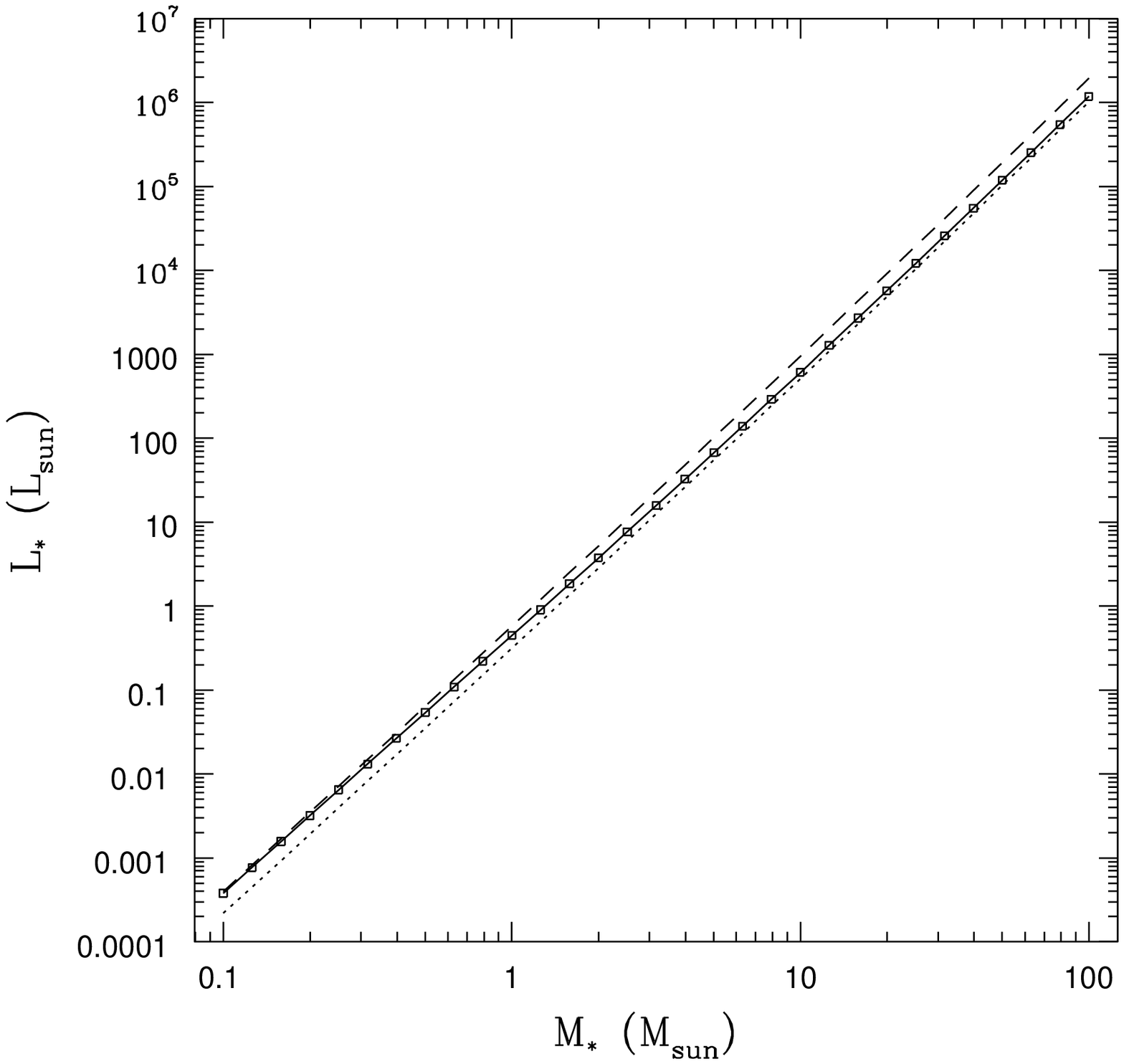}}
\caption{Stellar luminosity as a function of stellar mass for
standard values of the parameters. The three curves shown here
correspond to the $L_\ast-M_\ast$ relation for an $n$ = 3/2 polytrope
(dashed curve), an $n$ = 3 polytrope (dotted curve), and a model that
smoothly varies from $n$ = 3/2 at low masses to $n$ = 3 at high masses
(solid curve marked by symbols). All quantities are given in solar
units. }
\label{fig:lumvsmass} 
\end{figure}

For this simple polytropic stellar model, Figures \ref{fig:hrdiagram}
and \ref{fig:lumvsmass} show the H-R diagram and the corresponding
luminosity versus mass relation for stars on the zero age main
sequence (ZAMS). The three curves show different choices for the
polytropic indices: The dashed curves show results for $n=3/2$, the
value appropriate for low-mass stars.  The dotted curves show the
results for $n$ = 3, the value for high-mass stars. The solid line
(marked by symbols) show the results for $n$ varying smoothly between
$n = 3/2$ in the limit $M_\ast \to 0$ and $n = 3$ in the limit $M_\ast
\to \infty$.  We take this latter case as our standard model (although
the effects of changing the polytropic index $n$ are small compared to
the effects of changing the fundamental constants -- see \S 3).

One can compare these models with the results of more sophisticated
stellar structure models (\cite{kip, hansen}) or with observations of
stars on the ZAMS. In both of these comparison, this polytropic model
provides a good prediction for the stellar temperature as a function
of stellar mass. However, the luminosities of the highest mass stars
are somewhat low, mostly because the stellar radii from the models are
correspondingly low; this discrepancy, in turn, results from our
simplified treatment of nuclear reactions.  Nonetheless, this
polytropic model works rather well, and produces the correct stellar
characteristics $(L_\ast, R_\ast, T_\ast)$, within a factor of
$\sim2$, as a function of mass $M_\ast$, over a range in mass of
$\sim1000$ and a range in luminosity of $\sim 10^{10}$. This degree of
accuracy is sufficient for the purposes of this paper, and is quite
good given the simplifying assumptions used in order to obtain
analytic results. More sophisticated stellar models would include
varying values of $\conlum$ to incorporate more complex nuclear
reaction chains, detailed energy transport including convection, a
more refined treatment of opacity, and a fully self-consistent
determination of the density and pressure profiles (i.e., the
departures from our polytropic models). In particular, we can achieve
even better agreement between this stellar structure model and
observed stellar properties if we allow the nuclear reaction parameter
$\conlum$ to increase with stellar mass (as it does in high mass stars
due to the CNO cycle). In the spirit of this work, however, we use a
single value of $\conlum$, which corresponds to the case in which a
single nuclear species is available for fusion (this scenario thus
represents the simplest universes).

\section{Constraints on the Existence of Stars} 

Using the stellar structure model developed in the previous section,
we now explore the range of possible stellar masses in universes with
varying value of the stellar parameters. First, we find the minimum
stellar mass required for a star to overcome quantum mechanical
degeneracy pressure (\S 3.1) and then find the maximum stellar mass as
limited by radiation pressure (\S 3.2). These two limits are then
combined to find the allowed range of stellar masses, which can vanish
when the required nuclear burning temperatures becomes too high (\S
3.3). Another constraint on stellar parameters arises from the
requirement that stable nuclear burning configurations exist (\S 3.4).
We delineate (in \S 3.5) the range of parameters for which these two
considerations provide the limiting constraints on stellar masses and
then find the region of parameter space that allows the existence of
stars.  Finally, we consider the constraints implied by the Eddington
luminosity (\S 3.6) and show that they are comparable to those
considered in the previous subsections.

\subsection{Minimum Stellar Mass} 

The minimum mass of a star is determined by the onset of degeneracy
pressure. Specifically, for stars with sufficiently small masses,
degeneracy pressure enforces a maximum temperature which is below that
required for nuclear fusion. The central pressure at the center of a
star is given approximately by the expression 
\be 
\pcent \approx \left( {\pi \over 36} \right)^{1/3} 
G M_\ast^{2/3} \rhocen^{4/3} \, , 
\ee 
where the subscript denotes that the quantities are to be evaluated at
the center of the star.  This result follows directly from the
requirement of hydrostatic equilibrium (e.g., \cite{chandra}).  

At the low mass end of the range of possible stellar masses, the
pressure is determined by contributions from the ideal gas law and
from non-relativistic electron degeneracy pressure. As a result, the
central pressure of the star must also satisfy the relation
\be
\pcent = \left( {\rhocen \over \mion} \right) k \tcent + 
\konstant \left( {\rhocen \over \mion} \right)^{5/3} \, , 
\ee
where $\mion$ is the mean mass of the ions (so that $\rhocen / \mion$ 
determines the number density of ions) and where the constant 
$\konstant$ that determines degeneracy pressure is given by  
\be
\konstant = {\hbar^2 \over 5 m_e} \left( 3 \pi^2 \right)^{2/3} \, , 
\ee
where $m_e$ is the electron mass. Notice that we have also 
assumed that the star has neutral charge so that the number 
density of electrons is equal to that of the ions, and that 
$m_e \ll \mion$. 

Combining the two expressions for the central pressure and 
solving for the central temperature, we obtain 
\be
k \tcent = \left( {\pi \over 36} \right)^{1/3} 
G M_\ast^{2/3} \mion \rhocen^{1/3} - \konstant (\rhocen / \mion)^{2/3} 
\, . 
\ee
The above expression is a simple quadratic function of the 
variable $\rhocen^{1/3}$ and has a maximum for a particular 
value of the central density \cite{phil}, i.e., 
\be 
k T_{\rm max} = \left( {\pi \over 36} \right)^{2/3} 
{G^2 M_\ast^{4/3} \mion^{8/3} \over 4 \konstant} \, . 
\label{tmax} 
\ee 
If we set this value of the central temperature equal to 
the minimum required ignition temperature for a star, $\tig$, 
we obtain the minimum stellar mass
\be
M_{\ast {\rm min}} = \left( {36 \over \pi} \right)^{1/2}
{ \left( 4 \konstant k \tig \right)^{3/4} \over G^{3/2} \mion^{2} }
\, . 
\ee 
After rewriting the equation of state parameter $\konstant$ in 
terms of fundamental constants, this expression for the minimum 
stellar mass becomes 
\be 
M_{\ast {\rm min}} = 6 (3 \pi)^{1/2} \left( {4 \over 5} \right)^{3/4}
\left( {m_P \over \mion} \right)^2
\left( {k \tig \over m_e c^2} \right)^{3/4} M_0 \, . 
\label{massmin} 
\ee 
As expected, the minimum stellar mass is given by a dimensionless
expression times the fundamental stellar mass scale defined in
equation (\ref{mscale}). Notice also that the gravitational constant
$G$ enters into this mass expression with an exponent of --3/2, as
anticipated by equation (\ref{mscale}).
 
\subsection{Maximum Stellar Mass} 

A similar calculation gives the maximum possible stellar mass.  In
this case the central pressure also has two contributions, this time
from the ideal gas law and from radiation pressure $P_R$, where 
\be
P_R = {1 \over 3} a \tcent^4 \, , 
\ee 
where $a = \pi^2 k^4 / 15 (\hbar c)^3$ is the radiation constant.
Following standard convention \cite{phil}, we define the parameter
$\bcon$ to be the fraction of the central pressure provided by the
ideal gas law. As a result, the radiation pressure contribution is
given by $P_R = (1 - \bcon) \pcent$. The central temperature can be
eliminated in favor of $\bcon$ to obtain the expression
\be
\pcent = 
\left( {3 \over a} {(1 - \bcon) \over \bcon^4} \right)^{1/3} 
\left( {4 \rhocen \over \mbar} \right)^{4/3} \, , 
\ee
where $\mbar$ is the mean mass per particle of a massive star.  By
demanding that the star is in hydrostatic equilibrium, we obtain 
the following expression for the maximum mass of a star: 
\be
M_{\ast {\rm max}} = \left( {36 \over \pi} \right)^{1/2}
\left( {3 \over a} {(1 - \bcon) \over \bcon^4} \right)^{1/2} 
G^{-3/2} \left( {k \over \mbar} \right)^2 \, , 
\ee
which can also be written in terms of the fundamental mass 
scale $M_0$, i.e.,  
\be
M_{\ast {\rm max}} =  
\left( {18 \sqrt{5} \over \pi^{3/2} } \right) 
\left( {1 - \bcon \over \bcon^4} \right)^{1/2}  
\left( {m_P \over \mbar} \right)^2 \, M_0 \, , 
\label{maxmass}
\ee
where this expression must be evaluated at the maximum value of
$\bcon$ for which the star can remain stable. Although the requirement
of stability does not provide a perfectly well-defined threshold for
$\bcon$, the value $\bcon = 1/2$ is generally used \cite{phil} and
predicts maximum stellar masses in reasonable agreement with observed
stellar masses (for present-day stars in our universe). For this 
choice, the above expression becomes $M_{\ast {\rm max}} \approx 20 
(m_P/\mbar)^2 M_0$.  Since massive stars are highly ionized, 
$\mbar \approx 0.6 m_P$ under standard conditions, and hence 
$M_{\ast {\rm max}} \approx 56 M_0 \approx 100 M_\odot$ for our
universe. As shown below, this constraint is nearly the same as that
derived on the basis of the Eddington luminosity (\S 3.6).

\subsection{Constraints on the Range of Stellar Masses: \\ 
The Maximum Nuclear Ignition Temperature}   
 
As derived above, the minimum stellar mass can be written as a
dimensionless coefficient times the fundamental stellar mass scale
from equation (\ref{mscale}).  Further, the dimensionless coefficient
depends on the ratio of the nuclear ignition temperature to the
electron mass energy, i.e., $k \tig / m_e c^2$.  The maximum stellar
mass, also defined above, can be written as a second dimensionless
coefficient times the mass scale $M_0$.  This second coefficient
depends on the maximum radiation pressure fraction $\bcon$ and
(somewhat less sensitively) on the mean particle mass $\mbar$ of a
high mass star.  For completeness, we note that the Chandrasekhar mass
$\mchan$ \cite{chandra} can be written as yet another dimensionless
coefficient times this fundamental mass scale, i.e.,
\be
\mchan \approx {1\over 5} \, (2 \pi)^{3/2} \, 
\left( {Z \over A} \right)^2 M_0 \, , 
\label{chandra} 
\ee 
where $Z/A$ specifies the number of electrons per nucleon in the star. 

These results thus show that if the constants of the universe were
different, or if they are different in other universes (or different
in other parts of our universe), then the possible range of stellar
masses would change accordingly. We see immediately that if the
nuclear ignition temperature is too large, then the range of stellar
masses could vanish. If all other constants are held fixed, then the
requirement that the minimum stellar mass becomes as large as the
maximum stellar mass is given by
\be 
\left( {k \tig \over m_e c^2} \right) \ge {5 \over 4} 
\left( {360 \over 3 \pi^4} \right)^{2/3}
\left( {\sqrt{1 - \bcon} \over \sqrt{8} \bcon^2} \right)^{4/3} 
\left( {\mion \over \mbar} \right)^{8/3} \, \approx 1.4 
\left( {\mion \over \mbar} \right)^{8/3} \, , 
\label{maxburn} 
\ee 
where we have used $\bcon = 1/2$ to obtain the final equality. For
high mass stars in our universe, $\mbar / \mion = 0.6$, and the right
hand side of the equation is about 5.6. For the simplistic case where
$\mbar = m = \mion$, the right hand side is 1.4. In any case, this
value is of order unity and is not expected to vary substantially from
universe to universe.  As a result, the condition for the nuclear
burning temperature to be so high that no viable range of stellar
masses exists takes the form $k \tig/(m_e c^2) \simgreat 2$. For
standard values of the other parameters, the nuclear ignition
temperature (for Hydrogen fusion) would have to exceed $\tig \sim
10^{10}$ K.  For comparison, the usual Hydrogen burning temperature is
about $10^7$ K and the Helium burning temperature is about $2 \times
10^8$ K.  We stress that the Hydrogen burning temperature in our
universe is much smaller than the value required for no range of
stellar masses to exist --- in this sense, our universe is {\it not}
fine-tuned to have special values of the constants to allow the 
existence of stars.  The large value of nuclear ignition temperature
required to suppress the existence of stars roughly corresponds to the
temperature required for Silicon burning in massive stars (again, for
the standard values of the other parameters).  Finally we note that
the nuclear burning temperature $\tig$ depends on the fundamental
constants in a complicated manner; this issue is addressed below.

Equation (\ref{maxburn}) emphasizes several important issues.  First
we note that the existence of a viable range of stellar masses ---
according to this constraint --- does not depend on the gravitational
constant $G$. The value of $G$ determines the scale for the stellar
mass range, and the scale is proportional to $G^{-3/2} \sim
\alpha_G^{-3/2}$, but the coefficients that define both the minimum
stellar mass and the maximum stellar mass are independent of $G$. The
possible existence of stars in a given universe depends on having a
low enough nuclear ignition temperature, which requires the strong
nuclear force to be ``strong enough'' and/or the electromagnetic force
to be ``weak enough''. These requirements are taken up in \S 3.5.
Notice also that we have assumed $m_e \ll m_P$, so that electrons
provide the degeneracy pressure, but the ions provide the mass.

\subsection{Constraints on Stable Stellar Configurations} 

In this section we combine the results derived above to determine the
minimum temperature required for a star to operate through the burning
of nuclear fuel (for given values of the constants). For a given
minimum nuclear burning temperature $\tig$, equation (\ref{massmin})
defines the minimum mass necessary for fusion.  Alternatively, the
equation gives the maximum temperature that can be attained with a
star of a given mass in the face of degeneracy pressure. On the other
hand, equation (\ref{tcsolution}) specifies the central temperature
$\tcent$ necessary for a star to operate as a function of stellar
mass. We also note that the temperature $\tcent$ is an increasing
function of stellar mass.  By using the minimum mass from equation
(\ref{massmin}) to specify the mass in equation (\ref{tcsolution}), we
can eliminate the mass dependence and solve for the minimum value of
the nuclear ignition temperature $\tig$. The resulting temperature is
given in terms of $\thetacen$, which is given by the solution to the
following equation:
\be
\thetacen I(\thetacen) = \left( {2^{23} \pi^7 3^4 \over 5^{11} } \right) 
\left( {\hbar^3 \over c^2} \right) \left( {1 \over \beta \mzero^4} \right) 
\left( {1 \over m m_e^3} \right) \left( {G \over \kappa_0 \conlum} \right) \, . 
\label{iprofile}
\ee
Note that the parameters on the right hand side of the equation have
been grouped to include numbers, constants that set units,
dimensionless parameters of the polytropic solution, the relevant
particle masses, and the stellar parameters that depend on the
fundamental forces. Within the treatment of this paper, these latter
quantities could vary from universe to universe.  Notice also that we
have specialized to the case in which $\mbar = \mion = m$.

\begin{figure} 
\resizebox{\hsize}{!}{\includegraphics[]{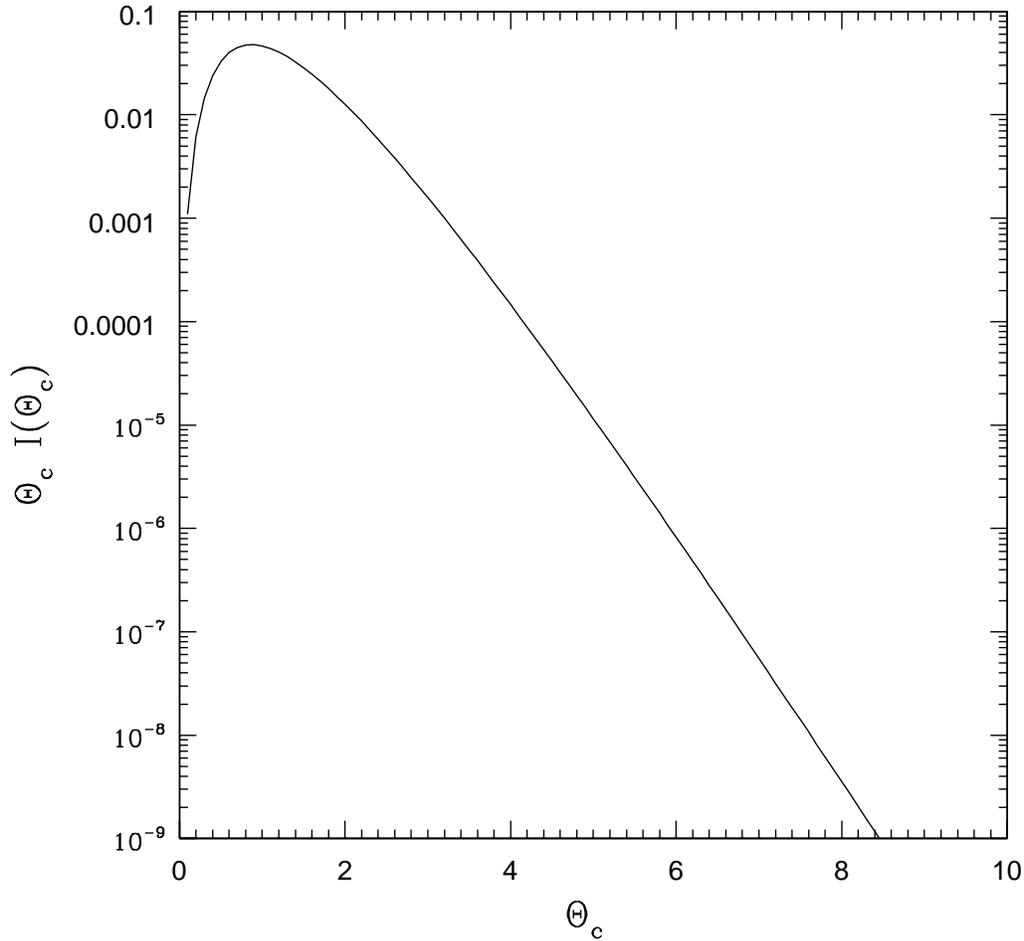}}
\caption{Profile of $\thetacen I(\thetacen)$ as a function of
$\thetacen = (E_G/4k \tcent)^{1/3}$. The integral $I(\thetacen)$
determines the stellar luminosity in dimensionless units and 
$\thetacen$ defines the central stellar temperature. This profile 
has a well-defined maximum near $\thetacen \approx 0.869$, where 
the peak of the profile defines a limit on the values of the
fundamental constants required for nuclear burning, and where the
location of the peak defines a maximum nuclear burning temperature
(see text). } 
\label{fig:iprofile} 
\end{figure}

The left hand side of equation (\ref{iprofile}) is determined for a
given polytropic index. Here we use the value $n=3/2$ corresponding to
both low-mass conventional stars and degenerate stars. The resulting
profile for $\thetacen I(\thetacen)$ is shown in Figure
\ref{fig:iprofile}. The right hand side of equation (\ref{iprofile})
depends on the fundamental constants and is thus specified for a given
universe.  In order for nuclear burning to take place, equation
(\ref{iprofile}) must have a solution --- the left hand side has a
maximum value, which places an upper bound on the parameters of the
right hand side. Through numerical evaluation, we find that this
maximum value is $\sim$0.0478 and occurs at $\thetacen \approx 0.869$.
The maximum possible nuclear burning temperature thus takes the form
\be 
\left( k T \right)_{max} \approx 0.38 E_G \, , 
\ee 
where $E_G$ is the Gamow energy appropriate for the given universe. 
The corresponding constraint on the stellar parameters required for 
nuclear burning can then be written in the form 
\be
{\hbar^3 G \over c^2 m m_e^3 \kappa_0 \conlum} \, \le 
{ 5^{11} \beta \mzero^4 \over 2^{23} \pi^7 3^4  } \, \, 
\Biggl[ \thetacen I(\thetacen) \Biggr]_{max} 
\approx 2.6 \times 10^{-5} \, , 
\label{conmax} 
\ee
where we have combined all dimensionless quantities on the right hand
side.  For typical stellar parameters in our universe, the left hand
side of the above equation has the value $\sim 2.4 \times 10^{-9}$,
smaller than the maximum by a factor of $\sim$11,000.  As a result,
the combination of constants derived here can take on a wide range of
values and still allow for the existence of nuclear burning stars. In
this sense, the presence of stars in our universe does not require
fine-tuning the constants.

Notice that for combinations of the constants that allow for nuclear
burning, equation (\ref{iprofile}) has two solutions. The relevant
physical solution is the one with larger $\thetacen$, which
corresponds to lower temperature. The second, high temperature
solution would lead to an unstable stellar configuration.  As a
consistency check, note that for the values of the constants in our
universe, the solution to equation (\ref{iprofile}) implies that
$\thetacen \approx 5.38$, which corresponds to a temperature of about
$9 \times 10^6$ K. This value is thus approximately correct: Detailed
stellar models show that the central temperature of the Sun is about
$15 \times 10^6$ K, and the lowest possible hydrogen burning
temperature is a few million degrees \cite{phil, kip, hansen}.

\subsection{Combining the Constraints} 

Thus far, we have derived two constraints on the range of stellar
structure parameters that allow for the existence of stars. The
requirement of stable nuclear burning configuration places an upper
limit on the nuclear burning temperature, which takes the approximate
form $kT \simless 0.38 E_G$.  In addition, the requirement that the
minimum stellar mass (due to degeneracy pressure) not exceed the
maximum stellar mass (due to radiation pressure) places a second upper
limit on the nuclear burning temperature, $k T \simless 2 m_e c^2$. As
a result, the reason for a universe failing to produce stars depends
on the size of the dimensionless parameter
\be
Q_F = 2 \alpha^2 {m \over m_e} \, , 
\ee
where $m$ is the mass of the nuclei that would experience reactions.
Note that $Q_F$ is proportional to the ratio of the Gamow energy to
the rest mass energy of the electron and has the value $Q_F \approx
0.2$ in our universe.  For $Q_F > 1$, stars can fail to exist due to
the range of allowed stellar masses shrinking to zero, whereas for
$Q_F < 1$ stars can fail to exist due to the absence of stable nuclear
burning configurations.

\begin{figure} 
\resizebox{\hsize}{!}{\includegraphics[]{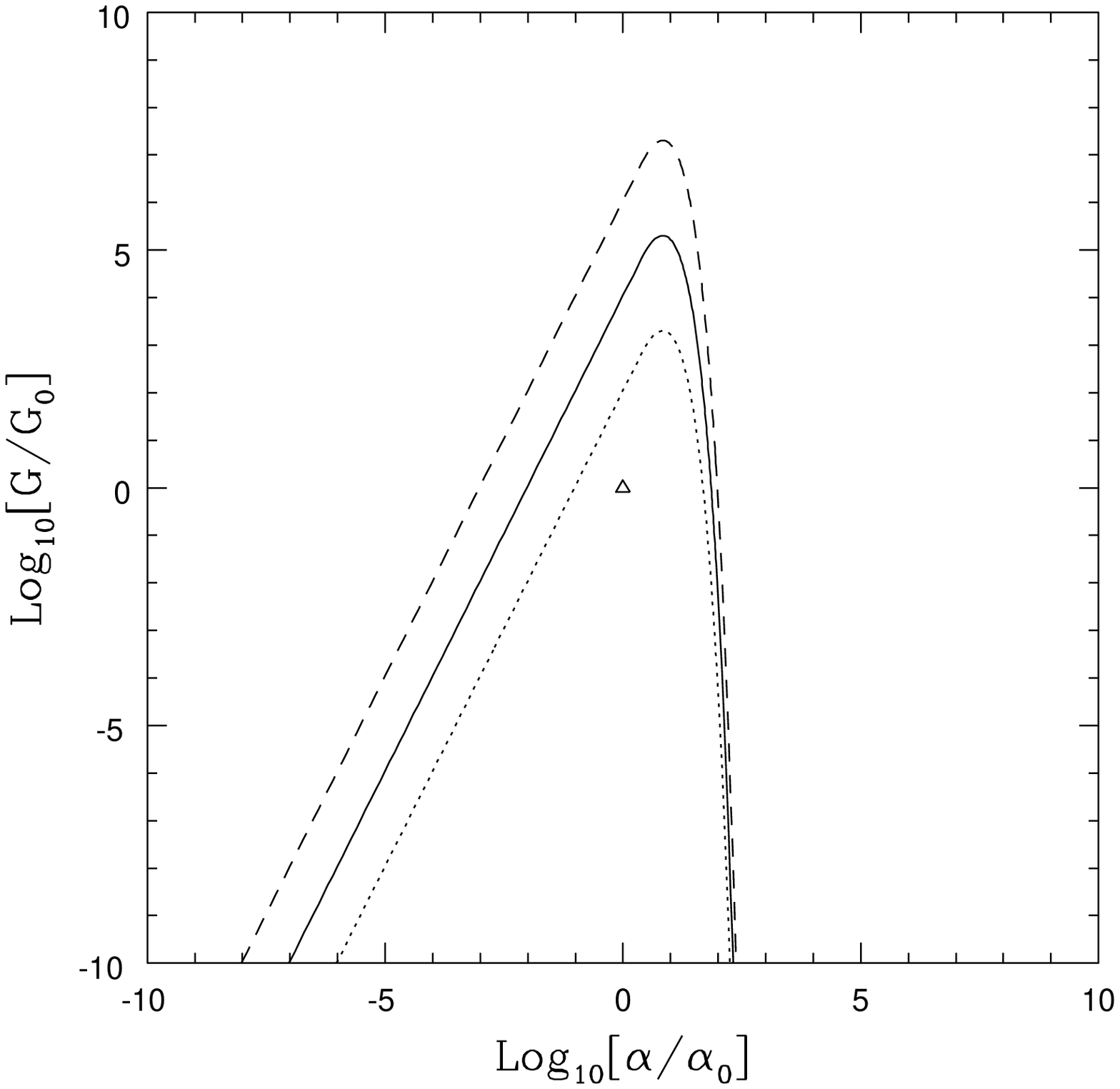}}
\caption{Allowed region of parameter space for the existence of
stars. Here the parameter space is the plane of the gravitational
constant $\log_{10}[G/G_0]$ versus the fine structure constant
$\log_{10}[\alpha/\alpha_0]$, where both quantities are scaled
relative to the values in our universe. The allowed region lies under
the curves, which are plotted here for three different values of the
nuclear burning constants $\conlum$: the standard value for $p$-$p$
burning in our universe (solid curve), 100 times the standard value
(dashed curve), and 0.01 times the standard value (dotted curve). 
The open triangular symbol marks the location of our universe in 
this parameter space. } 
\label{fig:agspace} 
\end{figure}

We can combine the constraints to delineate the portion of parameter
space that allows for the existence of stars.  For the sake of
definiteness, we fix the values of the particle masses and specialize
to the simplest case where the nuclear burning species has a single
mass $m$. We also assume that the stellar opacity scales according to
$\kappa_0 \propto \alpha^2$, as expected since $\kappa \sim \sigma_T /
m$ and $\sigma_T \propto \alpha^2$. With these restrictions, the
remaining stellar parameters that can be varied are the fine structure
constant $\alpha$, the gravitational constant $G$, and the nuclear
burning parameter $\conlum$. Note that $\alpha$ depends on the
strength of the electromagnetic force, $G$ depends on the strength of
gravity, and $\conlum$ depends on a combination of the weak and strong
nuclear forces, which jointly determine the nuclear reaction
properties for a given universe. Notice also that since we are fixing
particle masses, the gravitational constant $G$ is proportional to the
gravitational fine structure constant $\alpha_G$ (eq. [\ref{alphag}]).

Figure \ref{fig:agspace} shows the resulting allowed region of
parameter space for the existence of stars.  Here we are working in
the $(\alpha,G)$ plane, where we scale the parameters by their values
in our universe, and the results are presented on a logarithmic
scale. For a given nuclear burning constant $\conlum$, Figure
\ref{fig:agspace} shows the portion of the plane that allows for stars
to successfully achieve sustained nuclear reactions. Curves are given
for three values of $\conlum$: the value for $p$-$p$ burning in our
universe (solid curve), 100 times larger than this value (dashed
curve), and 100 times smaller (dotted curve).  The region of the
diagram that allows for the existence of stars is the area below the
curves.

Figure \ref{fig:agspace} provides an assessment of how ``fine-tuned''
the stellar parameters must be in order to support the existence of
stars.  First we note that our universe, with its location in this
parameter space marked by the open triangle, does not lie near the
boundary between universes with stars and those without. Specifically,
the values of $\alpha$, $G$, and/or $\conlum$ can change by more than
two orders of magnitude in any direction (and by larger factors in
some directions) and still allow for stars to function. This finding
can be stated another way: Within the parameter space shown, which
spans 10 orders of magnitude in both $\alpha$ and $G$, about one
fourth of the space supports the existence of stars.

Next we note that a relatively sharp boundary occurs in this parameter
space for large values of the fine structure constant, where $\alpha
\sim 200 \alpha_0$, and this boundary is nearly independent of the
nuclear burning constant $\conlum$. Strictly speaking, this
well-defined boundary is the result of the required value of $G$
becoming an exponentially decreasing function of $\alpha/\alpha_0$, as
shown in \S 3.7 below.  For the given range of $G$ and for values of
$\alpha$ above this threshold, the Gamow energy is much larger than
the rest mass energy of the electron, so that the maximum nuclear
burning temperature becomes a fixed value (that given by eq.
[\ref{maxburn}]), and hence the nuclear reaction rates are
exponentially suppressed by the electromagnetic barrier (\S 2.2).  On
the other side of the graph, for values of $\alpha$ smaller than
those in our universe, the range of allowed parameter space is limited
due to the absence of stable nuclear burning configurations (\S 3.4).
In this regime, for sufficiently large $G$, the nuclear burning
temperature becomes so large that the barrier disappears (and hence
stability is no longer possible). Since the nuclear burning
temperature $\tig$ required to support stars against gravity increases
as the gravitational constant $G$ increases, and since $\tig$ is
bounded from above, there is a maximum value of $G$ that can support
stars (for a given value of $\conlum$). For the value of $\conlum$
appropriate for $p$-$p$ burning in our universe, we thus find that
$G/G_0 \simless 2 \times 10^5$.  Finally, we note that ``stellar''
bodies outside the range of allowed parameter space can exist, in
principle, and can even generate energy, but they would not resemble
the stable, long-lived nuclear burning stars of our universe.

\subsection{The Eddington Luminosity} 

For a star of given mass, the maximum rate at which it can generate
energy is given by the Eddington luminosity. This luminosity defines a
minimum lifetime for stars. The Eddington luminosity can be written in
the form
\be
\lmax = 4 \pi c G M_\ast / \kapem \, , 
\label{eddington} 
\ee
where $\kapem$ is the opacity in the stellar photosphere.  For the
sake of definiteness, we take $\kapem$ to be the opacity appropriate
for pure electron scattering, which is applicable to hot plasmas where
the Eddington luminosity is relevant, i.e.,
\be
\kapem = {1 + X_1 \over 2} {\sigma_T \over m_P} \, , 
\ee
where $\sigma_T$ is the Thompson cross section and $X_1$ is the 
mass fraction of Hydrogen. Since the maximum luminosity implies a
minimum stellar lifetime, for a given efficiency $\epsilon$ of 
converting mass into energy, we obtain the following constraint 
on stellar lifetimes
\be 
t_\ast > \tstar = \epsilon \left( {1 + X_1 \over 3} \right) 
{\alpha^2 \over \alpha_G} {\hbar m_P \over (m_e c)^2} \, . 
\ee 
Since atomic time scales are given (approximately) by 
\be
t_{A} \approx {\hbar \over \alpha^2 m_e c^2} \, , 
\label{atomic} 
\ee 
the ratio of stellar time scales to atomic time scales is 
given by the following expression: 
\be
{\tstar \over t_{A} } = \epsilon \left( {1 + X_1 \over 3} \right) 
{\alpha^4 \over \alpha_G} {m_P \over m_e} \, , 
\ee 
where the expression has a numerical value of $\sim4 \times 10^{30}$ 
for the parameters in our universe. 

We can also use the Eddington luminosity to derive another upper limit
on the allowed stellar mass. Within the context of our model, the
stellar luminosity is given by equation (\ref{lstarsolve}). This
luminosity must be less than the Eddington luminosity given by
equation (\ref{eddington}), which implies a constraint of the form 
\be
{M_\ast \over M_0} \simless {4 \over \pi} \sqrt{60} \left( 
{\beta \mzero^3 \kappa_0 m_P \thetacen \over \sigma_T}   
\right)^{1/2} \, , 
\ee 
where we have specialized to the case of polytropic index $n=3$
(appropriate for high mass stars with large admixtures of radiation
pressure) and have taken $\mbar = m_P$. Note that since $\kappa_0 \sim
\sigma_T / m_P$, and since $\beta$ and $\mzero$ are given by the
polytropic solution (and are of order unity), the right hand side of
the above equation is approximately $50 \sqrt{\thetacen}$, as expected. 
In other words, the requirement that the stellar luminosity must be
less than the Eddington limit (eq. [\ref{eddington}]) produces nearly
the same bound on stellar masses as the requirement that the star not 
be dominated by radiation pressure (eq. [\ref{maxmass}]). 

Notice also that we expect $\kappa_0 \sim \sigma_T / m_P$ for other
universes, so that the general constraint takes the approximate form
$M_\ast / M_0 \simless 50 \sqrt{\thetacen}$. In addition, as shown by
Figure \ref{fig:iprofile}, the parameter $\thetacen$ is confined to a
narrow range --- the function $\thetacen I(\thetacen)$, and hence the
left hand side of equation (\ref{iprofile}), varies by 8 orders of
magnitude for $1 \simless \sqrt{\thetacen} \simless 3$.

\subsection{Limiting Forms} 

For much of the allowed parameter space where stars can operate; the 
value of $\thetacen$ is large compared to its minimum value;
specifically, this claim holds for the region of parameter space that
is not near the upper left boundary in Figure \ref{fig:agspace}. In
this case, we can derive an analytic asymptotic expression for the
integral function $I(\thetacen)$, which takes the form 
\be 
I (\thetacen) \sim 3 \thetacen {\rm e}^{- 3 \thetacen - 1}  
\left( {3 \pi \over \thetacen + 4/3} \right)^{1/2} \, \to 
(3/{\rm e}) \sqrt{3 \pi \thetacen} {\rm e}^{- 3 \thetacen} \, . 
\label{iasymp} 
\ee 
Comparing this asymptotic expression to the numerically determined
values, we find that equation (\ref{iasymp}) provides an estimate
that is within a factor of 2 of the correct result over the range 
$1 \le \thetacen \le 100$, where $I (\thetacen)$ varies by $\sim 128$
orders of magnitude.  

With this asymptotic expression in hand, we can find the relationship between 
the gravitational constant and the fine structure constant on the boundary of 
parameter space (as shown in Figure \ref{fig:agspace}). We find that 
\be
G \sim G_0 \, \exp \left[ - {3 \over 2} 
\left( {\alpha \over \alpha_0} \right)^{2/3} \right] \, . 
\ee 
At the edge of the allowed stellar parameter space, $G$ is thus an
exponentially decreasing function $\alpha$, which results in the
nearly vertical boundary shown in Figure \ref{fig:agspace}.

\section{Unconventional Stars} 

The results of the previous section show that stars can exist in a
relatively large fraction of the parameter space. On the other hand,
in order for stars to exist at all, the nuclear burning parameter
$\conlum$ must be nonzero; otherwise, stars, as objects powered by
nuclear reactions, cannot exist. In situations where $\conlum$ = 0, or
where the values of the other parameters conspire to disallow stars
(see Fig. \ref{fig:agspace}), other types of stellar objects could, in
principle, fill the role played by stars in our universe. This section
briefly explores this possibility with three examples: black holes
that generate energy through Hawking evaporation (\S 4.1), degenerate
dark matter stars that generate energy via annihilation (\S 4.2), and
degenerate baryonic matter stars that generate energy by capturing
dark matter particles which then annihilate (\S 4.3). We note that a
host of other possibilities exist (e.g., astrophysical objects powered
by proton decay), but a proper treatment of such cases is beyond the
scope of this present work.

\subsection{Black Holes} 

Black holes can exist in any universe with gravity and will generate
energy (at {\it some} rate) through Hawking evaporation (e.g.,
\cite{hawk}).  Further, the stellar structure of these objects depends
only on the gravitational constant $G$.  In order to consider black
holes playing the role of stars, however, we must invoke additional
constraints. For purposes of illustration, this section finds the
values of the fundamental constants for which black holes can serve as
stellar bodies to support ``life''. Specifically, in order for black
holes to fill the role played by stars in our universe, two
constraints must be satisfied: First, the black holes must live long
enough to allow for life to develop.  Second, the black holes must
provide enough power to run a biosphere.  The first constraint implies
that black holes must be sufficiently massive, whereas the second
constraint implies that the black holes must be sufficiently small.
The compromise between these two requirements provides an overall
constraint that must be met in order for black holes to play the role
of stars.

The lifetime of a black hole with mass $\mbh$ is given by 
\be
\tau_{bh} = {2650 \pi \over g_\ast \hbar c^4} G^2 \mbh^3 \, , 
\ee
where $g_\ast$ is the total number of effective degrees of freedom in
the radiation field produced through the Hawking effect. This lifetime
should be compared with the typical atomic time scale $\tau_A$ given
by equation (\ref{atomic}).  We thus have a constraint of the form
\be 
{\tau_{bh} \over \tau_A} = {2560 \pi \over g_\ast (\hbar c)^2} 
\alpha^2 G^2 \mbh^3 m_e \ge \nlife \, ,
\ee
where $\nlife$ is the number of atomic time scales required for life
to evolve. In our solar system, the number $\nlife \approx 10^{34}$,
which is also the number of atomic time scales in the life of a solar
type star.  Although the minimum value of $\nlife$ remains
uncertain, we expect it to be within a few orders of magnitude of this
value. For the sake of definiteness, we take the (somewhat optimistic) 
value of $\nlife = 10^{33}$ for this analysis. 

The second constraint is that the black hole must provide enough 
power to run a biosphere. In our solar system, the Earth intercepts 
about 100 quadrillion Watts of power from the Sun. We thus expect 
that the black hole must have a minimum luminosity and obey a 
constraint of the form 
\be
L_{\rm bh} = {g_\ast \hbar c^6 \over 7680 \pi} \left( G \mbh \right)^{-2} 
\ge \lmin \, , 
\ee 
where $\lmin$ is the minimum luminosity of a stellar object required
to support life. In general, this minimum value of luminosity will
vary with the values of the fundamental constants. In the absence of a
definitive theory, we adopt the following simple scaling law: The
energy levels $E_A$ of atoms vary according to $E_A \propto \alpha^2$,
and the atomic time scale varies as $t_A \propto \alpha^{-2}$.  In
order for the luminosity to provide the same number of atomic
reactions over the total lifetime of the system, the luminosity 
should scale with the fine structure constant as 
\be
\lmin = \lmin_0 (\alpha / \alpha_0)^4 \, , 
\label{lminscale} 
\ee
where $\lmin_0$ is the minimum necessary luminosity in our universe.
Although the value of this latter quantity is uncertain, we adopt
$\lmin_0 \approx 10^{17} {\rm erg} \, \, {\rm s}^{-1}$ as a representative
value. The scaling law of equation (\ref{lminscale}) is also not 
definitive, but rather illustrative. 

Combining the two constraints allows for the elimination of the mass, 
and thereby provides an overall constraint of the from 
\be
{\nlife \hbar^{1/2} (G/\alpha)^2 \over m_e} \le 
{c^7 \over 96 (15 \pi)^{1/2} \lmin^{3/2} } \, . 
\ee
If we scale this constraint using measured values of the constants, 
we obtain the relation
\be 
\left( {G \over G_0} \right) 
\left( {\alpha \over \alpha_0} \right)^{4} \le 
24 \left( {\nlife \over 10^{33} } \right)^{-1} 
\left( {\lmin_0 \over 10^{17} \, {\rm erg/s} } \right)^{-3/2} \, . 
\ee  
In our universe, black holes must have masses greater than about 
$6 \times 10^{13}$ g in order to last for $\nlife = 10^{33}$ atomic
time scales, and must have masses less than about $2 \times 10^{14}$ g
in order to produce enough power $(\lmin)$. As a result, a biosphere
could be powered by a black hole, although we have adopted somewhat
optimistic requirements, e.g., the required luminosity is only
$\lmin$, which is much less than a solar luminosity.  The largest 
obstacle, however, is the production of black holes with this mass
scale. 

\begin{figure} 
\resizebox{\hsize}{!}{\includegraphics[]{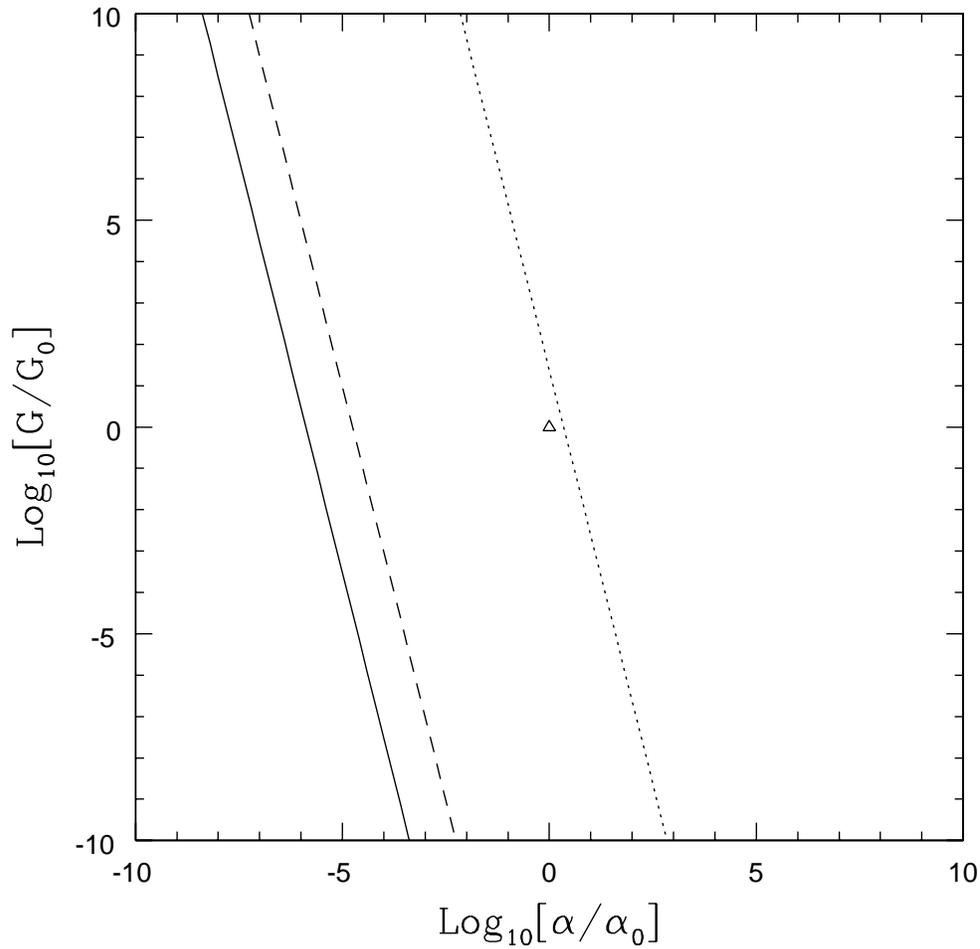}}
\caption{Allowed region of parameter space for the existence of
black holes that can play the role of stars. The parameter space is
the plane of the gravitational constant $\log_{10}[G/G_0]$ versus the
fine structure constant $\log_{10}[\alpha/\alpha_0]$, where both
quantities are scaled relative to the values in our universe. The
allowed region lies under the curves, which are plotted here for three
cases: The black hole luminosity is required to be greater than that
of the Sun (solid curve), a low-mass star (dashed curve), and the
solar luminosity intercepted by the Earth (dotted curve). The open 
triangle marks the location of our universe in this parameter space. }
\label{fig:bhspace} 
\end{figure}

Figure \ref{fig:bhspace} shows the region of parameter space for which
black holes can play the role of stars. To construct this diagram, we
assume that black holes must live $\nlife = 10^{33}$ atomic time
scales and produce enough luminosity. For this latter requirement, we
use the power intercepted from the Sun by the Earth (as a minimum
value; dotted curve), the luminosity of a low-mass star ($L \sim
10^{-3} L_\odot$; dashed curve), and 1.0 $L_\odot$ (solid curve), all
scaled according to equation (\ref{lminscale}).  If the black hole is
required to have luminosity in the stellar range, then the allowed
region of parameter space is highly constrained, in that the
parameters $(\alpha,G)$ must have values quite far from those in our
universe. In particular, the gravitational constant must be small (so
that the luminosity is large), and the fine structure constant must
also be small (so that atomic energy levels are low). If the necessary
luminosity is determined by $\lmin_0 = 10^{17}$ erg/s, however, black
holes can play the role of stars over a much wider range of parameter
space.
 
\subsection{Degenerate Dark Matter Stars} 

In principle, alternate universes can produce degenerate stars made of
dark matter particles. Such stars could exist in our universe as well,
although their formation is expected to be so highly suppressed that 
they play no significant role. This section considers the structure 
of these hypothetical objects in possible other universes. 

A degenerate star has the structure of an $n=3/2$ polytrope, 
with the constant $K$ in the equation of state given by 
\be
K = (3 \pi^2)^{2/3} {\hbar^2 \over 5 \mdm^{8/3} } \, , 
\ee 
where $\mdm$ is the mass of the dark matter particle.  Since the
constant $K$ is specified, we can solve directly for the stellar
properties. The mass-radius relation is given by
\be
M_\ast R_\ast^3 = \xi_\ast^3 \mzero {9 \pi^2 \over 128} 
\hbar^6 \mdm^{-8} G^{-3}  \, , 
\ee 
and the central density is given by 
\be
\rhocen = {32 \over 9 \pi^2 \mzero^2} 
{G^3 \mdm^8 M_\ast^2 \over \hbar^6} \, . 
\ee
For completeness, we note that the Chandrasekhar mass for 
these dark matter stars is given approximately by the expression 
\be
M_{ch} = \mu_0 { (3 \pi)^{1/2} \over 2} 
\left( {\hbar c \over G \mdm^2} \right)^{3/2} \mdm \, , 
\label{chandradm} 
\ee 
where $\mu_0 \approx 2.714$ for an $n = 3/2$ polytrope; this
expression does not include general relativistic corrections (e.g.,
\cite{st}). For reference, note that a typical expected value for the
dark matter particle mass, $\mdm = 100 m_P$, implies that this mass
scale $M_{ch} \approx 0.0007 M_\odot$.

For these stars, the luminosity is provided by annihilation 
of the dark matter particles. The annihilation rate per 
particle $\Gamma_1$ is given by 
\be
\Gamma_1 = n \langle \sigma_{\rm d} v \rangle \approx 
\sigma_{\rm d} \hbar n^{4/3} / \mdm \, ,
\ee
where $\sigma_{\rm d}$ is the cross section. To find the stellar 
luminosity due to dark matter annihilation, we must integrate 
over the star to find the total annihilation rate $\Gamma_T$: 
\be
\Gamma_T = {M_\ast \sigma_{\rm d} \hbar \over \mzero \mdm^2} 
n_{\rm c}^{4/3} \gamma_0 \, , \qquad {\rm where} 
\qquad \gamma_0 \equiv \int_0^{\xi_\ast} \xi^2 f^{7/2} d\xi \, . 
\ee
For this $n=3/2$ polytrope, $\gamma_0 \approx 0.913$. As a result, the
total annihilation rate is given by $\Gamma_T \approx N_T \Gamma_1/3$,
where $N_T$ is the total number of particles in the star and
$\Gamma_1$ is evaluated at the stellar center.  The corresponding
stellar luminosity is then given by
\be 
L_\ast = \left( {32 \over 9 \pi^2} \right)^{4/3} 
{\gamma_0 \over \mzero^{11/3} } {c^2 \over \hbar^7} 
\sigma_{\rm d} G^4 M_\ast^{11/3} \mdm^{25/3} \, . 
\ee
If the mass of the degenerate star were close to the Chandrasekhar
mass, the luminosity would be enormous, and its lifetime would be 
short (see below).  To put this in perspective, if we use reasonable 
values of the dark matter properties for our universe ($\mdm = 100
m_P$ and $\sigma_{\rm d}$ = $10^{-38}$ cm$^2$), then the mass required
to produce $L_\ast = 1.0 L_\odot$ is only about $M_\ast \sim 10^{-13}
M_\odot \sim 10^{20}$ g (about the mass of a large asteroid). As a
result, for the range of parameter space for which these objects play
the role of stars, the masses are far below the Chandrasekhar mass.

If the dark matter star starts its evolution with initial mass $M_0$
and later has a mass $M (t) \ll M_0$, then its age $\Delta t (M)$ is 
related to its current mass through the expression
\be
\Delta t (M) = {3 \over 8} \left( {9 \pi^2 \over 32} \right)^{4/3} 
{\mzero^{11/3} \over \gamma_0} {\hbar^7 \over \sigma_{\rm d}} 
G^{-4} M^{-8/3} \mdm^{-25/3} \, 
= {3 \over 8} {M c^2 \over L_{\ast} } \, , 
\label{dmtime} 
\ee
where $L_{\ast}$ is the luminosity of the star when it has mass $M$.
For example, if $M$ = $10^{20}$ g (the mass scale that generates
$L_\ast = 1.0 L_\odot$), the time scale from equation (\ref{dmtime})
is only about 100 days. In order for the time scale to be 1 Gyr, say,
the mass scale must be about $3 \times 10^{16}$ g, and the corresponding 
luminosity is only $\sim 10^{-13} L_\odot$= $4 \times 10^{20}$ erg/s, i.e., 
still substantially larger than the expected value of $\lmin_0$. 

When the masses are well below the Chandrasekhar mass (see above), the
star must satisfy two constraints. The first requirement is that the
star is sufficiently luminous, which implies that 
\be 
L_\ast = B {c^2 \sigma_{\rm d} G^4 \mdm^{25/3} \over \hbar^7} M^{11/3} 
\, \ge \lmin_0 (\alpha/\alpha_0)^4 \, , 
\ee
where we have defined a dimensionless constant $B$, 
\be 
B = \left( {32 \over 9 \pi^2} \right)^{4/3} {\gamma_0 \over \mzero^{11/3} } 
\approx 0.0060 \, . 
\ee
Next we require that the stellar lifetime is sufficiently long. 
In rough terms, this constraint can be written in the form 
\be
\Delta t (M) = {3 \over 8 B} 
{\hbar^7 \over \sigma_{\rm d} G^{4} \mdm^{25/3} }\, M_\ast^{-8/3} 
\, \ge {\hbar \nlife \over m_e c^2 \alpha^2} \, , 
\ee
where we have not made the distinction between $M$ and $M_\ast$ in
using equation (\ref{dmtime}). The first constraint puts a lower limit
on the mass, and the second constraint puts an upper limit on the mass. 
By requiring that both constraints be met simultaneously, the mass can 
be eliminated and a global constraint can be derived: 
\be
\left( {\alpha \over \alpha_0} \right)^{21/8} \, 
\left( {G \mdm^2 \over \hbar c} \right)^{3/2} \le \, 
C_B \, \, {m_e c^2 \over \lmin_0} \, \,  
\left( {m_e^{3} c^{10} \over \sigma_{\rm d}^3 \mdm \hbar^2 } \right)^{1/8} \, \, 
\left( {\alpha_0^2 \over \nlife } \right)^{11/8} \, , 
\label{dmlimit} 
\ee
where the constant $C_B$ = $(3/8)^{11/8} B^{-3/8} \approx 1.75$. This 
result defines the parameters necessary for dark matter stars to play
the role of ordinary stars (keep in mind that the formation of these
bodies remains a formidable obstacle).  The luminosity is determined
by the dark matter annihilation cross section, which is independent of
the constants that determine the physical structure of the star. As a
result, the parameter space of constants $(\alpha,G)$ considered here
always contains a region where these stars can operate: For fixed
properties of the dark matter ($\mdm$ and $\sigma_{\rm d}$), equation
(\ref{dmlimit}) delineates the portion of the $\alpha-G$ plane that
allows these degenerate dark matter objects to act as stars. On the 
other hand, one can use equation (\ref{dmlimit}) to constrain the
allowed dark matter properties for given values of $\alpha$ and $G$.

\subsection{Other Possibilities for Unconventional Stars} 

If the nuclear burning constant $\conlum$ = 0, then baryonic objects
can still, in principle, generate energy in a variety of ways.  In the
absence of nuclear reactions, stellar bodies will often tend to form
degenerate configurations, analogous to white dwarfs in our universe
(provided that their mass is below the relevant Chandrasehkar mass
scale). These degenerate objects can generate energy through several
channels, including residual heat left over from formation, proton
decay, and dark matter capture and annihilation.

In the latter case, dark matter particles are captured by scattering
off nuclei (which could be simply protons in a universe with no
nuclear reactions). After a scattering event, the recoil energy of the
dark matter particle can be less than the escape speed of the star,
and the particle can be captured. After capture, the dark matter
particles sink to the stellar center, where they collect until their
population is dense enough for annihilation to balance the incoming
supply of particles. The star thus reaches a steady state, where the
luminosity is given by the total capture rate. This process has been
discussed previously in a variety of context, including as a solution
to the solar neutrino problem \cite{press} and as a means to keep
white dwarfs hot beyond their cooling times \cite{al}.

The capture rate of dark matter particles is given by 
\be 
\Gamma = n_{\rm dm} \sigma_{\ast {\rm dm}} v_{rel} \, , 
\ee
where $n_{\rm dm}$ is the number density of dark matter particles,
$\sigma_{\ast {\rm dm}}$ is the total cross section for capture
subtended by the star, and $v_{rel}$ is the relative velocity. These
quantities depend on dynamical structure (distributions of density,
velocity, angular momentum) of the background halo of dark matter
\cite{bt87}. In our universe, for example, the capture rate of dark
matter particles by white dwarfs is of order $\Gamma \sim 10^{25}$
s$^{-1}$ \cite{al}.  With the capture rate specified, the
corresponding luminosity is given by
\be 
L_\ast = f_\nu \mdm \Gamma \, , 
\ee 
where $\mdm$ is the mass of the dark matter particles and where the
efficiency factor $f_\nu$ takes into account energy loss from the star
due to some fraction of the annihilation products being neutrinos.

In this scenario, the luminosity depends on the number density of dark
matter particles in the background (in the galactic halo in the
context of white dwarfs in our universe). This density is independent
of stellar properties. In a similar vein, the time scale over which
the luminosity can be maintained depends on the overall supply of dark
matter particles; this quantity is also independent of stellar
properties. Thus, for any values of the constants $(\alpha, G)$, 
considered here as the relevant parameters that specify stellar
properties, a universe {\it can} have the proper values of dark matter
densities and cross sections so that degenerate stars can serve in
place of nuclear burning stars. The specification of the allowed
parameter space depends on more global properties of the universe,
however, and is beyond the scope of this paper.

\section{Conclusion} 

In this paper, we have developed a simple stellar structure model (\S
2) to explore the possibility that stars can exist in universes with
different values for the fundamental parameters that determine stellar
properties. This paper focuses on the parameter space given by the
variables $(G, \alpha, \conlum)$, i.e., the gravitational constant,
the fine structure constant, and a composite parameter that determines
nuclear fusion rates. The main result of this work is a determination
of the region of this parameter space for which bona fide stars can
exist (\S 3). Roughly one fourth of this parameter space allows for
the existence of ``ordinary'' stars (see Figure \ref{fig:agspace}).
In this sense, we conclude that universes with stars are not
especially rare (contrary to previous claims), even if the fundamental
constants can vary substantially in other regions of space-time (e.g.,
other pocket universes in the multiverse).  Another way to view this
result is to note that the variables $(G, \alpha, \conlum)$ can vary
by orders of magnitude from their measured values and still allow for
the existence of stars.

For universes where no nuclear reactions are possible, we have shown
that unconventional stellar objects can fill the role played by stars
in our universe, i.e., the role of generating energy (\S 4).  For
example, if the gravitational constant $G$ and the fine structure
constant $\alpha$ are smaller than their usual values, black holes can
provide viable energy sources (Figure \ref{fig:bhspace}).  In fact,
all universes can support the existence of stars, provided that the
definition of a star is interpreted broadly. For example, degenerate
stellar objects, such as white dwarfs and neutron stars, are supported
by degeneracy pressure, which requires only that quantum mechanics is
operational. Although such stars do not experience thermonuclear
fusion, they often have energy sources, including dark matter capture
and annihilation, residual cooling, pycnonuclear reactions, and proton
decay. Dark matter particles can also (in principle) form degenerate
stellar objects (see \S 4).

In order to assess the suitability of non-nuclear power sources, one
must specify how much power is required, and for how long.  In this
work we have used the power that Earth intercepts from the Sun as the
minimum benchmark value $\lmin_0$, and scaled the necessary power
according to equation (\ref{lminscale}) to account for variations in
the fine structure constant; similarly, the required amount of time is
taken to to $\sim1$ Gyr, scaled by the atomic time of equation
(\ref{atomic}).  These choices are not definitive, and hence
alternative scalings can be explored.

The issue of alternate values for the fundamental constants, as
considered herein, is related to the issue of time variations in the
constants in our universe. However, current experiments place rather
strong limits on smooth time variations, with time scales exceeding
the current age of the universe (see the review of \cite{uzan}).
Another possibility is for the constants to have different values at
other spatial locations within our universe, although this scenario is
also highly constrained \cite{barrow}.

This paper has focused on stellar structure properties. An important
related question (beyond the scope of this work) is whether or not
stellar bodies can be readily made in universes with varying values of
the constants. Even if the laws of physics allow for stellar objects
to exist and actively burn nuclear fuel, there is no guarantee that
such bodies will be produced in significant numbers. In our universe,
for example, there is a moderate mismatch between the mass range of
possible stars and the distribution of masses of stellar bodies
produced by the star formation process.  At the present cosmological
epoch, star formation produces objects over the entire possible range
of stellar masses, with additional bodies produced in the substellar
range (brown dwarfs).  The matching is relatively good, in that the
fraction of bodies in the brown dwarf range is small, only about 1 out
of 5 \cite{luhman}.  Since the masses of these objects are small, the
fraction of the total mass locked up in the smallest bodies is even
smaller, less than 5 percent. On the other hand, nearly all of the
stars in our universe have small masses. As one benchmark, only about
3 or 4 out of a thousand stars are larger than the $\sim 8 M_\odot$
threshold required for stars to experience a supernova explosion,
whereas stellar masses can extend up to $\sim100 M_\odot$. The high
mass end of the possible mass range is thus sparsely populated.  The
corresponding match between the range of allowed stellar masses and
the mass range of objects produced can be quite different in other
universes.

In future work, another issue to be considered is coupling the effects
of alternate values of the fundamental constants to the cosmic
expansion, big bang nucleosynthesis, and structure formation. Each of
these issues should be explored in the same level of detail as stellar
structure is studied in this work. With the resulting understanding of
these processes, the coupling between them should then be determined.

Finally, we note that this paper has focused on the question of
whether or not stars can exist in universe with alternate values of
the relevant parameters. An important and more global question is
whether or not these universes could also support life of some
kind. Of course, such questions are made difficult by our current lack
of an {\it a priori} theory of life. Nonetheless, some basic
requirements can be identified (with reasonable certainty): In
addition to energy sources (provided by stars), there will be
additional constraints to provide the right mix of chemical elements
(e.g., carbon in our universe) and a universal solvent (e.g., water).
These additional requirements will place additional constraints on
the allowed region(s) of parameter space.

\medskip 

Acknowledgment: We thank Greg Laughlin for useful discussions. This
work was supported by the Foundational Questions Institute through
Grant RFP1-06-1 and by the Michigan Center for Theoretical Physics.

\end{document}